\documentstyle[12pt]{article}
\input epsf
\def\bea{\begin{eqnarray}}
\def\eea{\end{eqnarray}}
\newcommand \Pomeron {I\!\!P}

\begin{document}
\title{Signals for 
 black body limit  in coherent
ultraperipheral heavy ion collisions. }
\author{
L. Frankfurt\\
\it School of Physics and Astronomy, Raymond and Beverly Sackler\\
\it Faculty of Exact Science, Tel Aviv University, Ramat Aviv 69978,\\
\it Tel Aviv , Israel\\
M. Strikman\\
\it Pennsylvania State University, University Park, Pennsylvania 16802\\
M. Zhalov\\
\it Petersburg Nuclear Physics Institute, Gatchina 188350, Russia}
\date{}
\maketitle
\centerline {\bf ABSTRACT}
We argue that study of total cross section of photoabsorption and coherent
 photoproduction of $\rho,\rho^{\prime}$-mesons in ultraperipheral
heavy ion collisions (UPC) is effective method to probe onset of black body
limit(BBL) in the soft and hard QCD interactions. We illustrate the
expected features of the onset of BBL using generalized vector dominance
model. We show that this model describes very well
$\rho$-meson coherent photoproduction at $6 \leq E_{\gamma} \leq 10 GeV$.
In the case of $\rho$-meson production we find a UPC cross section 
which is a
factor $\sim 1.5 $ larger than the one found by 
Klein and Nystrand.
The advantages of the process of coherent dijet production to probe
onset of  BBL in hard scattering regime where decomposition over 
the twists becomes
inapplicable are explained and relative importance of the
$\gamma +Pomeron$ and $\gamma +\gamma$ mechanisms is estimated.

\section{Introduction}

Studies of the coherent interactions of photons with nucleons and
nuclei were one of the highlights of the strong interaction studies of
the seventies, for the excellent  summaries
see \cite{yenn,Shaw}.

The fundamental  question which one can investigate in the coherent
 processes is
how interactions change for
different types of projectiles with increase of the size/thickness
of the target. Several regimes appear possible. In the case of a hadronic
projectile (proton, pion, etc)  high-energy interactions with the nucleus
rather rapidly approach a black body limit (BBL)
in which the total cross section
of the interaction is equal to $2\pi R_A^2$. Another extreme limit is the
interaction of small size projectiles (or wave packages). In this case
at sufficiently high energies the system remains
frozen during the passage
through  the nucleus and the regime of color transparency is reached
in which the amplitude of interactions is  proportional to the gluon density
of the nucleus which is somewhat smaller than the sum of the
nucleon  gluon  densities due to the leading twist nuclear shadowing.
 In this regime the cross section of interaction
rapidly grows with energy reflecting the fast increase of the gluon
densities at small $x$ and large $Q^2$
and it may reach ultimately the black limit of interaction
from the perturbative domain.  (This limit can correspond to quite
different perturbative 
QCD dynamics, in particular it could be reached already at 
 $x\geq 10^{-3}$ where  $\ln x $ effects are a small correction.)
The BBL for the interaction of the small size dipoles 
 with heavy nuclei represents a new regime of interactions 
 when 
the leading twist approximation and therefore
the whole notion of the parton distributions becomes inapplicable for the
description of hard QCD processes in the small $x$ regime.
Obviously there should exist
also many cases when the projectile
 represents a superposition of configurations
of different sizes (leading to fluctuations of the strength of interaction).

In this respect interactions of photons with heavy nuclei provide unique
opportunities since the photon wave function
 contains both the hadron-like configurations
(vector meson dominance) and the direct photon configurations
 (small $q\bar q$ components).
The important advantage of the  photon is that
at high energies the BBL is manifested in diffraction into a multitude
of the hadronic final states  (elastic diffraction $\gamma \to  \gamma $
is negligible)
while in the hadron case only  elastic diffraction survives  in the BBL
and details of the dynamics leading to this regime remain hidden.
Spectacular manifestations of BBL in (virtual) photon diffraction include
strong enhancement of the large mass tail of
the diffractive spectrum as compared to the expectations of the
the triple Pomeron limit,  large cross section of the
high $p_t$ dijet production \cite{BBL}.

In preQCD time V.Gribov explored the  complete absorption of hadrons by
heavy nucleus to calculate the  total cross section of
photo(electro)production processes off heavy nuclei through the hadron
polarization operator for the photon $\rho(M^2)$:
\begin{equation}
F_T (x,Q^2)=\int \limits^{2q_0/R_A}_{m_0^2} dM^2 {2\pi R_A^2\over 12 \pi^3}
{Q^2M^2\rho(M^2)\over (M^2+Q^2)^2},
\label{gr}
\end{equation}
where $q_0=\omega_{\gamma}$ is the photon energy, 
$m_0^2\approx m_{\rho}^2$.
The upper cutoff in the integral in the black body limit formulae
(the Gribov approximation) comes from the nucleus form factor
\begin{equation}
-t_{min} R_A^2/3\approx \left({M^2+Q^2\over 2q_0}\right)^2 R_A^2/3  << 1.
\end{equation}
The distinctive feature of Eq.(\ref{gr}) is that the contribution of large
masses in the wave function of projectile photon (a direct photon
contribution)  is not suppressed.
Consequently, 
Eq. (\ref{gr}) leads to 
$\sigma_{\gamma A}^{tot}\propto  2\pi R_A^2\alpha_{em} 
\ln (2q_0/R_Am_{\rho}^2)$ 
for $A\gg 1$ (this is
 qualitatively
different from the hadron case where 
 $\sigma_{hA}^{tot} \approx  2\pi R_A^2$), and grossly violates 
expectations of the Bjorken scaling
for the $Q^2$ dependence of $\sigma_{\gamma A}^{tot}$.

To overcome this puzzle  J.Bjorken suggested a long 
time ago
the aligned jet model in which only 
$q\bar q$ pairs with small $p_t$  can interact while
high $p_t $ configurations in the photon wave function remain sterile
\cite{bj}. Existence of sterile states has been explained later
as due to the color transparency phenomenon \cite{FS88}.
More recently it was  understood that  states which behave as  sterile at
moderate energies, may interact at high enough energies with a hadron
target with  cross sections comparable to that for soft QCD phenomena.

Thus the Gribov's assumptions are justified in QCD for the interaction
 of a range of  hadronic components of the photon wave
function with heavy nucleus target.  At the same time 
even at small $x$ some  components 
are still small 
enough, so that they  interact with a small cross section
- for these components the color transparency still survives. Hence
one needs smaller $x$ to reach the BBL than allowed by the cutoff in the
integral in Eq.(\ref{gr}). This  $x$-range
 was not reached so far experimentally in $ep$
collisions.

It is worth emphasizing that
 the hypothesis of BBL corresponds to the assumption that at 
sufficiently small 
$x$ partons with large virtuality interact 
with heavy nuclei without any suppression
with a cross sections $ \approx  2\pi R_A^2$. 
It is this feature of the BBL which is responsible
for the gross violation of the Bjorken scaling and 
for the above mentioned qualitative difference of the energy dependence of 
 $\sigma_{tot}^{\gamma A} $ and  $\sigma_{tot}^{h A}$.

One of striking features of the BBL regime is the suppression of nondiagonal
transitions in the photon interaction with heavy nuclei \cite {Gribov}
\footnote{In Ref.\cite{Brodsky} it was assumed that one can neglect
interference effects for a nucleon  target also.
In this case in order to preserve the Bjorken scaling one has to make an
 assumption that the cross section of the interaction of  heavy mass
configurations with nucleons decreases $\propto 1/M^2$.
}.
Indeed in the BBL  the dominant contribution to the coherent  diffraction
originates from ``a shadow'' of the fully absorptive interactions
at impact parameters $b\leq R_A$ and hence the orthogonality condition 
is applicable.

Very little is known experimentally so far about coherent photon induced
diffractive phenomena
due to
the problems of separating events where nucleus remained intact
in the fixed target experiments and absence of  electron-nucleus colliders.
New opportunities for the investigation 
of photon-nucleus
 interactions become available in
ultraperipheral collisions (UPC) of heavy nuclei at RHIC and LHC. These
studies will allow to extract the cross section of the coherent
$\gamma A $ interactions up to  $\sqrt{s} \sim 60 (15)$  GeV (LHC/RHIC)
due to a possibility to select the events where colliding nuclei
remain intact or nearly intact, see e.g.
\cite{Klein97,FELIX}, see Refs.\cite{baur,Baur2002} for the recent
reviews and extensive lists of references.
Recently we investigated possibilities of studying 
color transparency and perturbative color opacity related to the
leading twist gluon shadowing in $J/\psi$ UPC and commented on 
the onset of BBL
for $J/\psi$ production \cite{FSZpsi}.

In this paper we will 
continue studies of the UPC phenomena.
Our aim is to evaluate pattern of soft QCD phenomena in the
proximity to black body limit, disappearance of color transparency
phenomenon in the hard processes with increase of energies. 
We will study photoproduction of $\rho$-mesons
and the $I=1$ mesonic states with masses $1.5 \leq M^2\leq 4 GeV^2$ usually
generically referred to as a $\rho^{\prime}$-meson
in the processes: $\gamma + A \to V +A, ~ A+ A \to A + A + V; 
V=\rho,\rho^{\prime}$.
 To
visualize expected new phenomena we will use generalized vector dominance
model which takes into account fluctuations of the interaction
strength to show that relative yield of $\rho$ and $\rho^{\prime}$ mesons is
sensitive to the onset of BBL physics in soft regime.
We will argue that  the  production of two jets in the  process
$A+A\to A +A + 2~jets$ in collisions of  heavy nuclei provides a new 
effective method of  probing the  onset of BBL for the  hard QCD phenomena.

\section{Vector meson production off nuclei in\\
 the generalized vector dominance model}

In this section we will use  generalized vector dominance 
model to describe coherent photoproduction of 
hadronic states of $M\leq 2$ GeV off nuclei.

The vector dominance model (VDM) \cite{Sakurai} was first suggested as
an explanation of the nuclear shadowing in the interactions of photons
with nuclei \cite{Stodolsky} in a close connection with the Bell discussion
of the shadowing in neutrino - nucleus scattering \cite{Bell}.
It was also pointed out in \cite{Stodolsky} that at sufficiently
high energies heavier states may become important.
Importance of extending VDM to include the heavy
mass states - Generalized VDM (GVDM) 
was further emphasized and explored  in the late sixties
\cite{Gribov, Brodsky}.
In particular one needs large mass states to explain the slope of  $Q^2$
dependence of structure functions at small $Q^2$ -${1/(1+Q^2/0.71)}$
behavior instead of ${1/(Q^2+m_{\rho}^2)}^2$ predicted by  the VDM.

The main ambiguity in such an extension was the  issue of nondiagonal
transitions where a photon initially converts into one vector state - $V_1$
which through diffractive interactions with a nucleon converts into
another state $V_2$. Such amplitude would interfere with the process of
direct production of $V_2$. Such nondiagonal transitions were introduced 
in a number of GVD models \cite{GVDM,GVDM1}. Physically the
 importance of such transitions could be justified on the basis of the 
interpretation of the early Bjorken scaling for moderately small $x\sim 10^{-2}$
as due to the color transparency phenomenon - presence 
in the
virtual photon of hadron type and point-like type
configurations  \cite{FS88}.  Presence of nondiagonal 
transitions is  also crucial for ensuring a 
quantitative matching with perturbative QCD  regime
 for $Q^2\leq$ ~few GeV$^2$\cite{FGS97}.
Hence it is reasonable to use GVDM  for  the modeling of the
production of the light states off nuclei.

The amplitude of the vector meson production off a nucleon
can be written within the GVDM  as
\begin{equation}
A(\gamma + N\to V_j + N )=\sum_{i} {e\over f_{V_i}}A(V_i + N \to V_j + N),
\label{GVDM}
\end{equation}
where $f_{V_i}$ are connected to the width of decay of the corresponding
resonance in the process $e^+e^-\to hadrons$.
In the case of nuclei calculation of the amplitude of the
Glauber scattering with production of a meson $V$
requires taking into  account both  the nondiagonal transitions
due to the transition of the photon to a different meson $V'$ in the
vertex $\gamma \to V' $ and due to change of the meson in
multiple rescatterings like $\gamma \to V \to V' \to V$.
This physics is equivalent to inelastic shadowing phenomenon familiar from
hadron-nucleus scattering \cite{Gribovinel}.
The Glauber model for the description of these processes is well known,
so,  we present here only the basic formulae which we will use
to calculate the photoproduction cross section\footnote{In this calculation
we neglect the triple Pomeron 
contribution which is present at high energies.
This contribution though noticeable for the scattering off the lightest nuclei
becomes a very small correction for 
 the scattering of heavy nuclei due
the strongly absorptive 
nature of interaction at the central impact parameters.}
\begin{equation}
{\frac {d\sigma_{\gamma A\to VA}(t)} {dt}} =
\pi {\left|  \int \limits_{0}^{\infty} J_{0}({p_t}b)\Gamma (b) bdb\right|}^2
\label{glatot}
\end{equation}
Here $J_{0}(z)$ is the Bessel function,
 $p_t=\sqrt {t_{min}-t}$, $-t_{min}=\frac {M_V^4} {4q_0^2}$ is longitudinal
momentum transfer in $\gamma -V$ transition, and $\Gamma ({\vec b})$ 
is the nuclear profile function
 which is obtained in impact parameter space from the solution of the coupled
multichannel Glauber equations for production of vector mesons 
$\rho,\rho^{\prime}$ which  takes into account the finite coherence length
 effects due to the longitudinal
momentum transfers
(see e.g.  \cite{Pautz:qm} for the explicit expressions).

In Ref. \cite{Pautz:qm} the simplest nondiagonal model
(which is a truncation of a more general model \cite{GVDM,GVDM1})
was considered with two states $\rho$ and $\rho^{\prime}$ which
have the same diagonal amplitudes of scattering off a nucleon
and the fixed ratio of coupling constants
\begin{equation}
f_{\rho^{\prime}}/f_{\rho}=\sqrt{3},
\label{primconst}
\end{equation}
while the ratio of the nondiagonal and diagonal amplitudes
\begin{equation}
{\frac {A(\rho + N \to \rho^{\prime} +N)} {A(\rho + N \to \rho +N)}}=-\epsilon,
\label{mixpar}
\end{equation}
and the value $\sigma^{tot}_{\rho N}$ were found from the fit to
the
forward
$\gamma +A\to \rho +A$ cross sections
measured at $\omega_{\gamma}=$6.1, 6.6 and 8.8 GeV\cite{MIT}.
It was pointed out that this model
with reasonable values of $\sigma_{\rho N}$ and $\epsilon$ 
 allows to bring the value of
$f_{\rho}$ determined from the photoproduction of $\rho$-mesons off protons
assuming approximate equality of the cross sections
 of $\rho-N$ and $\pi -N $ interactions
into a good agreement with the $e^+e^-$ data thus removing
a long standing 20\% discrepancy between two determinations.
One should emphasize here that in the logic of GVDM  $\rho^{\prime}$-meson
approximates the
hadron production in the interval of hadron masses $\Delta M^2 \sim 2 GeV^2$,
so the values of the production cross section
refer to the corresponding
mass interval.

\begin{figure}
    \begin{center}
        \leavevmode
        \epsfxsize=1.\hsize
        \epsfbox{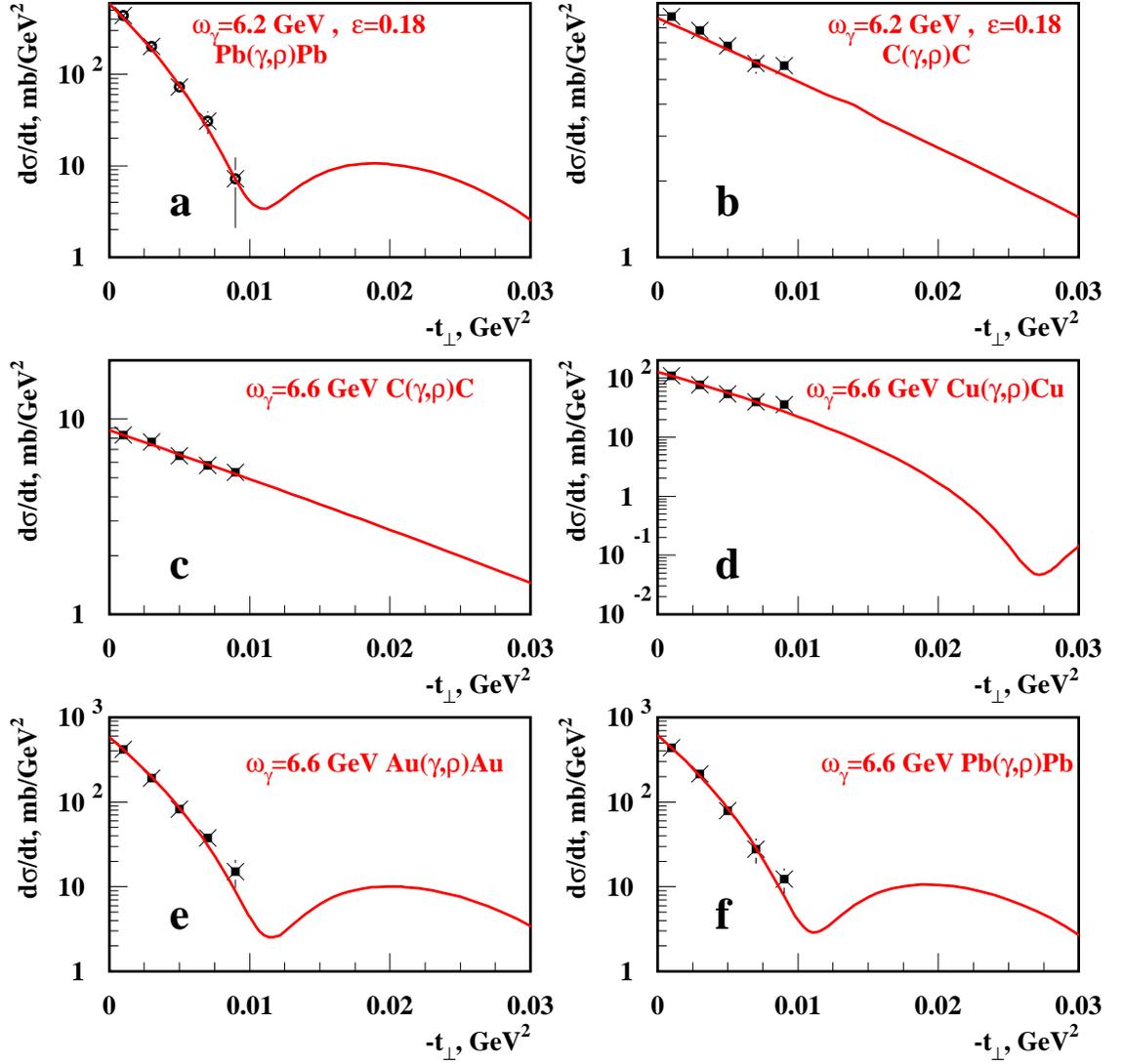}
    \end{center}
\caption{Description of the $\rho$-production data
\cite{MIT} by the GVDM Glauber
model with the value $\epsilon$ =0.18.}
\label{anadis}
\end{figure}

As a first step we shall refine the model and then compare it with
more detailed experimental data. First of all we diminish the
dependence on the nuclear structure parameters by
calculating the nuclear densities
in the Hartree-Fock-Skyrme  (HFS) approach. This model
not only provided an excellent
description (with an accuracy $\approx 2\%$)
of the nuclear root mean square radii and
the binding energies of spherical
nuclei along the periodical table from carbon to uranium\cite{HFS}
but also was successfully used to describe in the Glauber
approximation such detailed
characteristics of the nuclear structure as the shell model
momentum distributions in the high energy (p,2p)\cite{p2p}
and (e,e'p)\cite{eep} reactions. Next, we fixed
the values of the total
cross section of the $\rho N$ interaction and
$\eta_{\rho N}=\frac {\Re e A_{\rho N}} {\Im m A_{\rho N}}$
using the corresponding parameterizations suggested
in the Landshoff-Donnachie model\cite{LD}.
Accounting for the nondiagonal $\rho-\rho^{\prime}$ transitions
the value of $\epsilon $ was looked for to provide a best fit to
the differential cross section of
the $\rho$-meson photoproduction off lead at $\omega_{\gamma}=6.2$ GeV and
 $p_t^2=0.001$ $GeV^2$.
As a result  (Fig.\ref{anadis}a) we found $\epsilon=0.18$
which  is indeed very close to
 the lower end of the range $\epsilon=0.2 \div 0.28$ suggested in \cite{Pautz:qm}.
 Note that this value
leads to a suppression of the differential
cross section of the $\rho$-photoproduction in  $\gamma +p\to \rho +p$
by a factor of
$(1-\epsilon/\sqrt{3})^2\approx 0.80$
 practically coinciding with
phenomenological  renormalization factor ${ R=0.84}$ introduced in \cite{LD}
to achieve the best fit of the elementary $\rho$-meson photoproduction
forward cross section in the VDM which neglects mixing effects.

With all parameters fixed we calculated the  differential
cross sections of $\rho$-production off nuclei and found a good agreement
(Fig.\ref{anadis}b-f) with available data\cite{MIT}.

\begin{figure}
    \begin{center}
        \leavevmode
        \epsfxsize=0.5\hsize
        \epsfbox{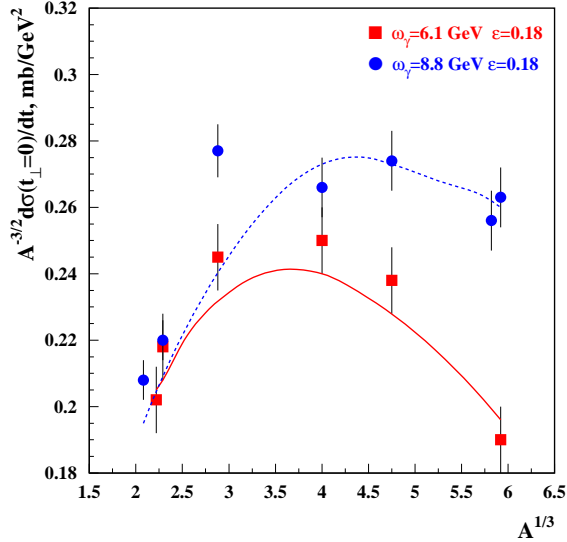}
    \end{center}
\caption{Description of the A-dependence of 
forward $\rho$-production data
\cite{MIT} by the GVDM Glauber
model with $\epsilon$ =0.18.}
\label{anadepf}
\end{figure}
A rather small systematic  discrepancy with the data at
$p^2_t \approx 0.01\,GeV^2$ appears to be due to 
 the incoherent $\rho$ photoproduction which is strongly
suppressed for the very small $p_t$ but gives 
a contribution comparable to the coherent one for 
$p^2_t \approx 0.01\,GeV^2$. 

We  have also checked the description of the A-dependence for the forward
$\rho$ photoproduction cross section (Fig.\ref{anadepf}).  
In difference from  Ref.\cite{LD} we did not find any evidence for an 
 increase 
of $\epsilon$ by almost  $50\%$ 
(from 0.2 to 0.28) when the energy of photons is increased from 6.2 GeV up 
to 8.8 GeV.

\begin{figure}
    \begin{center}
        \leavevmode
        \epsfxsize=.5\hsize
        \epsfbox{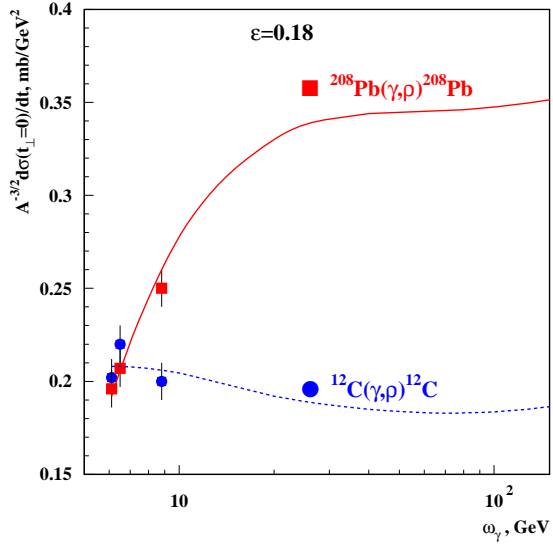}
    \end{center}
\caption{The energy dependence of the $\rho$-photoproduction cross section
calculated in the GVDM+Glauber model.}
\label{epend}
\end{figure}

\begin{figure}
    \begin{center}
        \leavevmode
        \epsfxsize=.5\hsize
        \epsfbox{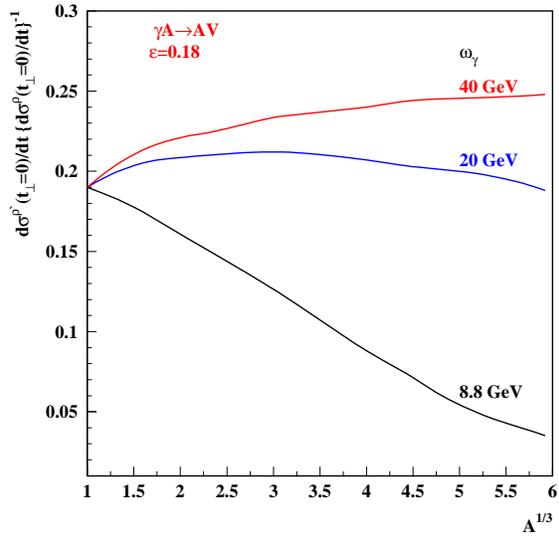}
    \end{center}
\caption{The A-dependence of the ratio of ${\rho^\prime}\over {\rho}$
forward photoproduction cross sections
calculated in the GVDM+Glauber model.}
\label{erat}
\end{figure}

 As far as we know previously this important check of the Glauber
model predictions in the vector meson production off $A > 2$ nuclei has 
never  been performed in  such  self-consistent way.
In view of a good agreement of the model with the data
on $\rho$-meson production in the low energy domain we will use this model
to consider the $\rho$ meson photoproduction at higher energies of
photons. The increase of the coherence length  with the photon energy
 leads to a  qualitative difference in the energy dependence of
the coherent vector meson production off light
and heavy nuclei (Fig. \ref{epend})  and to a  change of the A-dependence
for the ratio of the forward $\rho^{\prime}$ and $\rho$-meson production
 cross sections between $\omega_{\gamma} \sim 10 GeV$ 
and $\omega_{\gamma}\sim 50 GeV$ (Fig.\ref{erat}).  
The observed pattern reflects the
difference of  the 
coherence lengths of the $\rho$-meson and a heavier 
$\rho^\prime$-meson
 which is important for the intermediate
photon energies $\leq 30 GeV$

 Unfortunately no experimental data are available at the moment
on the  coherent $\rho^{\prime}$ photoproduction  and on
the $\rho$ photoproduction at energies $\geq$ 10 GeV.  Such studies maybe
possible with the HERMES detector at DESY and in the E-160  experiment
at SLAC. On the other hand, a very promising way to collect such data
would be a study of the coherent light vector meson production in the
ultraperipheral ion collisions (UPC) at RHIC and LHC
where one can explore the wide range of the quasi-real photon energies.

\section{Vector meson production in ultraperipheral collisions}
Production of vector mesons in ultraperipheral heavy ion collisions can be
expressed in the Weizsacker - Williams (WW)
approximation through the cross section
of the  vector meson production in $\gamma A$ scattering

\begin{equation}
{\frac {d\sigma_{AA\to AAV}} {dy}}=
2\int d{\vec b}T_{AA}({\vec b})n(y,{\vec b})
\sigma_{\gamma A\to VA}(y).
\label{rapdist}
\end{equation}
Here $y$ is rapidity of the produced vector meson,
$T_{AA}({\vec b})$ is the thickness function of
colliding nuclei on the impact parameter ${\vec b}$,
$n({\vec b},y)$ is the flux of photon with energy
$w=\frac {m_{V}} {2} e^{y}$ emitted by one of nuclei and
$\sigma_{\gamma A\to VA}(y)$ we calculated  integrating the
 Eq.(\ref{glatot}) over the momentum transfer
in the range $t_{min}\leq t \leq \infty$.

\begin{figure}
\begin{center}
        \epsfxsize=.6\hsize
        \epsfbox{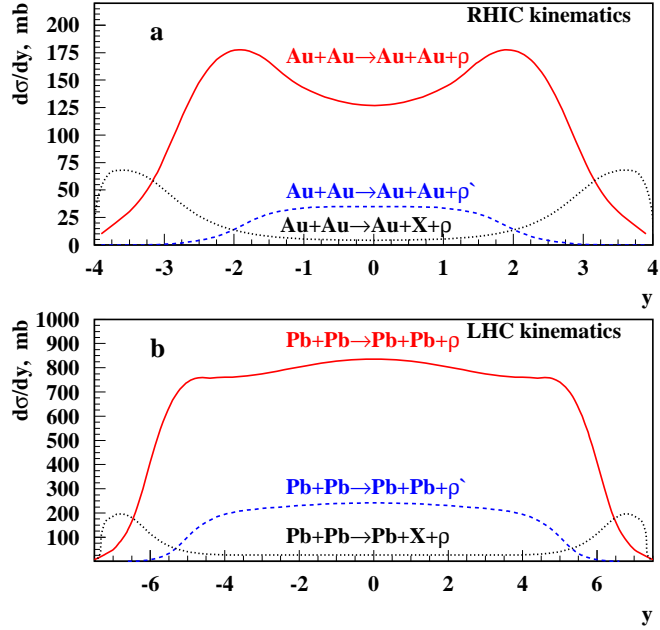}
\caption{Rapidity distributions for the light vector meson
production at RHIC and LHC}
\end{center}
\label{rapid}
\end{figure}

As we discussed in Section 2,  the GVDM with 
the value of $f_{\rho}$ fixed to the value determined from the 
$e^+e^-$  annihilation
gives a better description of the  cross section of the
coherent $\rho $ production from nucleons. We also demonstrated that
it gives a very good description of the absolute cross section and
$t$-dependence of the cross section of the $\rho$-meson photoproduction off
nuclei. Hence it is natural to expect that it would  provide a 
 reliable predictions for production of vector mesons in  UPC.
In particular,  we calculated within this model coherent cross sections of
both the $\rho$ and $\rho^{\prime}$ mesons.
The inelastic diffractive contribution is expected to be rejected using 
 the veto from Zero Degree Neutron Calorimeter which is
implemented  in the RHIC experiments and is planned for the  LHC.
 This veto is the  least effective for the single inelastic diffraction
 as this process 
will often result in the  events where one nuclear proton is removed 
and the residual nucleus remains in the ground or a low excitation
state.
Our calculation of the single inelastic diffraction shown in Fig.\ref{rapid}
demonstrates that this background is very   small 
for a wide range of central rapidities.

\begin{figure}
    \begin{center}
        \leavevmode
        \epsfxsize=.5\hsize
        \epsfbox{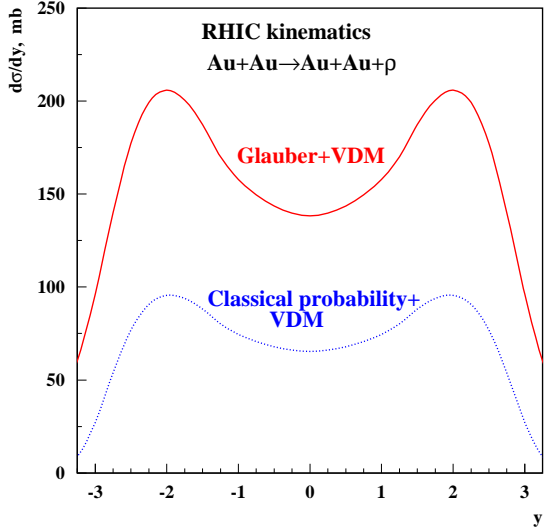}
    \end{center}
\caption{Comparison of the rapidity distributions calculated in VDM+
classical mechanics formula for total cross section(dotted line)
 with calculations within VDM+Glauber model(solid line)}
\label{klefig}
\end{figure}

The results of our calculations for the total cross sections
are given in Table \ref{tcrsec}.
\begin{table}
\begin{center}
\begin{tabular}{|c|c|c|}\hline
                   &      AuAu at RHIC  &    PbPb  at LHC    \\ \hline\hline
coherent $\rho$    &     934 mb         &   9538 mb          \\ \hline
coherent $\rho^{\prime}$   &     133 mb         &   2216 mb          \\ \hline
incoherent $\rho$  &     201 mb         &    846 mb          \\ \hline
\end{tabular}
\end{center}
\caption{Total cross sections of $\rho$ and $\rho^{\prime}$ production
 in UPC at RHIC and LHC.}
\label{tcrsec}
\end{table}
It should be emphasized that we have got the cross sections of the coherent
$\rho$ production considerably larger than estimates  in
 Ref. \cite{Klein991} where the
first quantitative study of the coherent $\rho$-meson production
in kinematics of the peripheral ion collisions at RHIC and LHC was
presented. In \cite{Klein991} as well as in \cite{Klein992}
the cross section was calculated as:
\begin{equation}
d \sigma_{\gamma +A \to V +A}=
\frac {\alpha_{em}} {4{f_{\rho}^2}}{\sigma_{tot}^2}(\rho A)
\int \limits_{t_{min}}^{\infty} dt  F_A^2(t),
\label{klein}
\end{equation}
where $F_A(t)$ is the  nuclear form factor.
Further it was assumed in \cite{Klein991} that $\sigma_{tot}(\rho A)$ is
given by the classical mechanics formula:
\begin{equation}
\sigma_{tot}(\rho A)=\int d^2\vec{b}[1-\exp(-\sigma_{tot}(\rho N)T(\vec{b}))],
\label{classic}
\end{equation}
where $T(\vec{b})$ is the usual thickness function.
It's easy to estimate that this formula leads to a
substantially smaller  value of the total cross
section than the quantum mechanical Glauber expression
$\sigma_{tot}(\rho A)=2\int d^2\vec{b}
[1-\exp(-{{\sigma_{tot}^{\rho N}}\over {2}}T(\vec{b}))]
$
  -
 a factor of two smaller for heavy enough nuclei:
$\sigma_{tot}(\rho A)=\pi R_A^2$ instead of $2 \pi R_A^2$. To show
explicitly the difference in results we compare  in Fig.~\ref{klefig} the
rapidity distributions obtained  in the VDM+Glauber model  with correct
accounting  for the longitudinal momentum transfer but without nondiagonal
terms (solid line) and result of calculations(dashed line) with the same
parameters(\cite{LD}) and the HFS nuclear form factor but in the model
based on Eqs. (\ref{klein},\ref{classic}) used in \cite{Klein991}.

In the follow up paper \cite{Klein992}
authors considered $p_t$ distribution of the
produced  vector mesons and made an interesting observation that the
amplitudes of the production of a vector mesons produced when a left moving
nucleus emits the photon and when right moving nucleus emits a photon should
destructively interfere. Due to the condition that 
essential impact
parameters in AA collisions are larger than $2R_A$ a significant interference
occurs only for $p_t\leq 1/2R_A $ corresponding to
$p_t\leq$ 10 MeV \cite{Klein992}.
This $p_t$ range constitutes a small fraction  of whole
permitted phase volume and hence the interference effects
can be neglected for the case of the
cross sections integrated over $p_t$ which were required to
calculate the rapidity distributions presented here (Fig.\ref{rapid}).

In the case of $\rho$ production corrections due to nondiagonal
transitions are relatively  small ($\sim$ 15\%)
 for the case of scattering off a nuclei. As a
result we find that the GVD cross section is close
to the one calculated in the VD model for heavy nuclei as well.

Situation is much more interesting for
 $\rho^{\prime}$ production.  In this case cross section
of production of $\rho^{\prime}$ off a nucleon
is strongly suppressed as compared to the  case when the $\rho
\leftrightarrow  \rho^{\prime}$  transitions are switched off.
The extra suppression factor is $\approx 0.5$.

In accordance with the general argument of Gribov the non-diagonal
transitions disappear in the limit of large A
(black body limit)  due to the condition of
orthogonality of hadronic wave functions \cite{Gribov}.
Hence we expect that in the limit of $A\to \infty$:
\begin{equation}
{d\sigma (\gamma + A \to V_1 +A )/dt
\over d\sigma (\gamma + A \to V_2 +A )/dt}_{\left|A\to
\infty\right.}
=\left(f_2/f_1\right)^2.
\label{bbl}
\end{equation}
In reality the $\rho$-meson is a broad resonance
which also interferes with the nonresonance $\pi^+\pi^-$ continuum,
and $\rho^{\prime}$ represents a set of overlapping
resonances and continuum. Also the detectors are likely to
be able to detect only some of the final states.
Hence it is convenient to use
a more general relation for the productions of
states $h_1, h_2$ of invariant masses $M_1^2,M_2^2$:
\begin{equation}
{d\sigma (\gamma + A \to h_1 +A )/dt \over
d\sigma (\gamma + A \to h_2 +A )/dt}_{\left|A\to \infty\right.}
={\sigma(e^+e^-\to h_1)\over \sigma(e^+e^-\to h_2)}.
\label{bbl1}
\end{equation}

\begin{figure}
    \begin{center}
%        \leavevmode
        \epsfxsize=1.\hsize
        \epsfbox{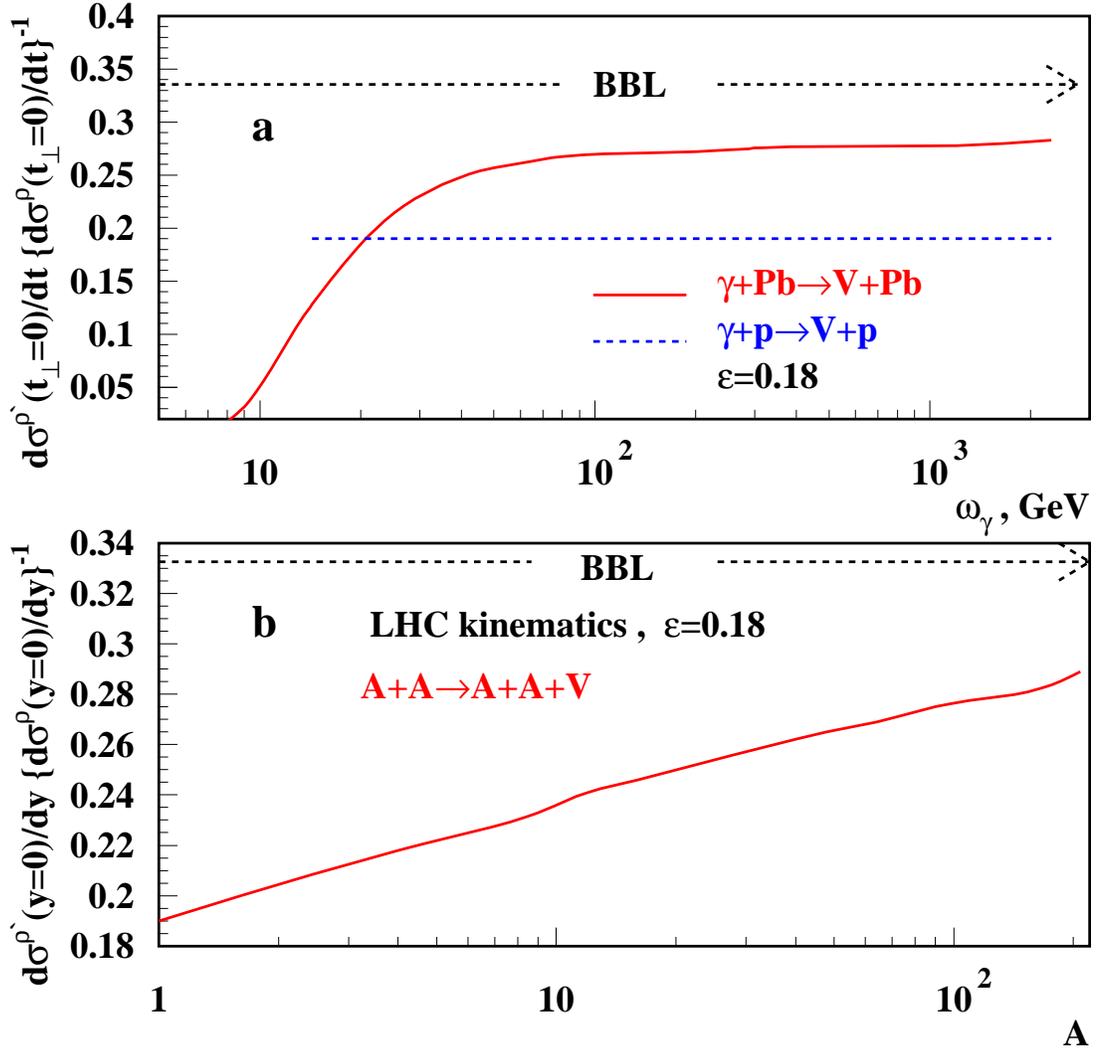}
    \end{center}
\caption{a. Energy dependence of the ratio of $\rho^{\prime}$
and $\rho$-meson production
cross sections, b. A-dependence of the ratio  of $\rho^{\prime}$ 
and $\rho$-meson
 production forward cross sections in kinematics at LHC.  }
\label{bblfig}
\end{figure}

Indeed we have found from calculations that in the case of the
coherent photoproduction off lead the nondiagonal transitions becomes strongly
suppressed with increase of the photon energy.
As  a result  the $\rho^{\prime}/\rho$ ratio increases,
exceeds the ratio of the $\gamma p\to Vp$
forward cross sections calculated with accounting
for $\rho-\rho^{\prime}$ transitions already
at $\omega_{\gamma}\geq 50$ GeV and becomes close to  the value
of $f^2_{\rho}/f^2_{\rho^{\prime}}$  which can be
considered as the limit when one can treat the interaction with
the heavy nucleus as a  black one. The same trend
to BBL is seen from A-dependence presented for kinematics
at LHC corresponding the value of
energy $W_{\gamma p}=60$ GeV (Fig.\ref{bblfig}) .

It is worth noting here that presence of nondiagonal transitions
which in terms of
the formalism of the 
scattering eigen states \cite{GW} corresponds to the
fluctuations of the interaction cross section leads to a substantial
modification of the  pattern of the approach to BBL. For example if one
would neglect nondiagonal transitions
one would
have to reduce both $\rho -N $ and $\rho^{\prime} - N $ cross sections in order
to keep the values of the production cross sections 
in $\gamma +p\to \rho +p$
the same as in the 
considered  GVDM.
For the $\rho$-meson  the reduction effect is 
a small 
correction $(1- \frac {\epsilon} {\sqrt 3}) \approx 0.9$, while 
the cross section of $\rho^{\prime}-N $
 interaction is  reduced by a substantially
larger
factor ${(1 - \sqrt 3 \epsilon)} \approx 0.7$ 
This would lead to a noticeable
 reduction of the total cross section 
of the $\rho^{\prime}-A$ interaction as compared to
the BBL value of $2\pi R_A^2$ and reduces 
the $\rho^{\prime}/\rho$ ratio   for $A\sim 200$ by $\approx 10\%$  
as  compared to reduction by a factor 0.9
in the original model. At the same time in a number of GVD models
 it is assumed that 
$\sigma_{tot}(VN)\propto 1/M^2$. 
In such a model the $\rho^{\prime}/\rho$ ratio for $Pb$ 
would be reduced by  a factor $\sim 3$.

The general BBL expression 
for the differential cross section of the production of the invariant
mass $M^2$ \cite{BBL} is

\begin{equation}
{{d\sigma_{(\gamma +A\to ``M''+A)}}
\over dt dM^2} ={\alpha_{em} \over 3 \pi}{(2\pi R_A^2)^2\over 16\pi}
{\rho(M^2)\over M^2} {4\left|J_1(\sqrt{-t}R_A)\right|^2
\over -t R_A^2}.
\label{ccsb}
\end{equation}

Hence by comparing the extracted cross section of the diffractive
production of states with certain masses with the
black body limit result - Eq.(\ref{ccsb}) one would be able to
determine up to what masses in the photon wave function
interaction remains black.
Onset of BBL limit for hard processes should reveal itself also in the
faster increase with energy of cross sections of photoproduction of
excited states with that for ground state meson.
It would be especially advantageous for these studies
 to use a set of nuclei - one medium range like $Ca$ and
another heavy one - one could remove the edge effects and use the
length of about 10 fm of nuclear matter.

Note in passing that an interesting change of the low-mass dipion
spectrum is expected in the discussed limit. It should be strongly
suppressed as compared to the the case of scattering
off proton where nonresonance continuum is much larger
than in $e^+e^-\to \pi^+\pi^-$ process.

\section{Diffractive dijet production}

For the $\gamma A$ energies which will be available at LHC
one may expect  that the BBL  in the scattering off heavy
nuclei would be a good approximation 
for the masses $M$ in the photon wave function  
up to  few GeV. This is 
the domain which is described by perturbative QCD for
 $x\sim 10^{-3}$ for the  proton
targets
and  larger $x$ 
for scattering off nuclei. The condition of large longitudinal 
distances - small longitudinal
transfer will be applicable in this case up to quite large values of the
produced diffractive mass (though it will not hold for masses
above 3 GeV or so at RHIC). Really $x_{eff}= M^2/s_{\gamma N}= M/2E_N$ will
be $\sim 10^{-3}$ for $M=4 GeV$ for  $y=0$. So that the condition
$l_{coh}=1/m_N x_{eff}\gg 2 R_A$ is satisfied.

In the BBL the dominant channel of diffraction for large masses is
production of two jets with the total cross section    given
by Eq.(\ref{ccsb}) and with a characteristic angular distribution
$(1+ \cos^2 \theta)$, where $\theta$ is the 
c.m. angle 
\cite{BBL}. On the contrary in the perturbative
QCD limit the
diffractive dijet production except charmed jet production
is strongly suppressed \cite{brodsky,diehl}. The suppression
is due to the structure of the coupling of the wave function of the
real photon wave
 to two gluons when calculated in the lowest order in $\alpha_s$.
   As a result  in the
real photon case hard diffraction involving light quarks is connected to
production of $q\bar q g$ and higher states.
Thus the dijet photoproduction should be very sensitive to the
onset of BBL regime.

Note that in the case of photon nucleon scattering at
$\omega_{\gamma}\sim 100 GeV$ \cite{Chapin}
the normalized differential 
${\frac {1} {\sigma_{tot}^{\gamma N}}}{d \sigma/dM^2}$ 
for diffraction into large masses ($\ge 2 GeV$)
is very similar to that for the pion nucleon scattering and appears
to be dominated by the triple Reggeon limit corresponding to the
process where a photon first converts to a $\rho$ meson and next a large
mass is produced in the $\rho-N$ diffractive scattering.
Since the triple Pomeron coupling constant is quite small this process should be
a small correction in the BBL. Besides in this limit the triple Pomeron
 process is
screened by the multiple Pomeron exchanges and originates solely
from the scattering off the rim of the nucleus. Hence it is suppressed
at least by a factor $\sim A^{1/3}$ as compared to the process of direct
diffraction into heavy masses.

A competing process for photoproduction of dijets off heavy nuclei is
production of dijets in $\gamma -\gamma$ collisions where the second photon
is provided by the Coulomb field of the nucleus. 
Note that the dijets produced in this process have positive
C-parity and hence this amplitude does not  interfere
with the amplitude of the dijet production in
the $\gamma \Pomeron$ interaction which have negative C-parity.

For the calculation of the cross section of dijet production in
$\gamma +\gamma$ collisions we use the lowest order
perturbative 
QCD result which coincides  up to the
number of colors factor and summation over the  quark flavors
with the well known
QED result for the lepton pair production in $\gamma \gamma$ collisions:
\begin{equation}
{d\sigma (\gamma + \gamma\to jet +jet)
\over d \Omega }= 3\sum_i e_{q_i}^4
\alpha_{em}^2{1\over M^2}\left[{2\over \sin^2\theta}-1\right].
\label{dijet2}
\end{equation}
Here the sum over the quark flavors goes over quarks with $m_q\ll
M/2$ and $p_t^{jet}$ is sufficiently large to suppress non-perturbative
contribution.
Using the  Weizsacker - Williams approximation we evaluate the
ratio of the  $\gamma \gamma$ and $\gamma \Pomeron $
contributions to the dijet production in AA collisions in the BBL
with the logarithmic accuracy:
\begin{equation}
R={{{d\sigma_{\gamma \gamma}}(A+A\to dijet+A+A)}
\over
{{d\sigma_{\gamma \Pomeron}}(A+A\to dijet +A+A)}}=
{\frac {\sum_i e_{q_i}^4} {\sum_i e_{q_i}^2}}
\frac {16Z^2 \alpha^2_{em} }
{{M^2}R^2_A sin^2\theta}\ln {{2q_0} \over {M^2 R_A}}.
\label{ratio}
\end{equation}
In the derivation of Eq. (\ref{ratio}) we
 neglected a  difference of  the
energy dependences of the processes.
For the kinematics of interest (large $p_t$ of jets and region of produced
masses $M \le \,3\,GeV$)  
$\theta=90^o$ in the center of mass of the produced system
 and   we can account for three lightest
flavors, hence
${\frac {\sum_i e_{q_i}^4} {\sum_i e_{q_i}^2}}=1/3$.
One can easily see that  $R \ll 1$
 for production of high $p_t$ jets corresponding to 
$\sin \theta \sim 1$, and hence the 
$\gamma \gamma$ contribution can be safely neglected.

It is worth emphasizing that at the energies below the BBL where
diffraction of the photon to dijets can be legitimately calculated in the
lowest  order in $\alpha_s$
(cf. calculation   of a similar process of dijet production in 
the pion - hadron scattering in Ref. \cite{FMS}) the 
 electromagnetic mechanism is much more
important. It is enhanced by a factor  $1/\alpha_s^2$ and becomes much
more prominent with increase of $p_t$ of the jet.
Also it it enhanced for very small total momentum of the dijet system.
 Observation of the last  effect  is hardly feasible, cf. 
the above discussion of the vector meson production.

\section{Conclusions}

We demonstrated that ultraperipheral AA collisions is effective method
of probing onset of BBL regime in hard processes at small $x$.
We have demonstrated that the Glauber model
predicts a significantly larger coherent $\rho$-meson
production rates than the previous calculations.
We predict a significant  increase of the
ratio of the yields of $\rho,\rho^{\prime}$
mesons in coherent processes off heavy nuclei due to the blackening of
the soft QCD interactions in which fluctuations of the interaction strength
are present. An account of nondiagonal transitions leads to a prediction of
a significant enhancement of production of 
heavier diffractive states especially production of high $p_t$
dijets. Study of these channels
may allow to get an important information on the onset of the black body
limit in the diffraction of real photons.

We thank J.Bjorken, S.Brodsky, G.Shaw for  useful discussions and GIF,
 CRDF and DOE for support.

\end{document}